\documentclass[useAMS]{tMPH2e}

\usepackage{amssymb}
\usepackage{amssymb}
\usepackage{amsmath}
\usepackage{graphicx}
\usepackage{amsbsy}
\usepackage{natbib}

\received{}

\title{Phase behavior of polydisperse sticky hard spheres: \\
analytical solutions and perturbation theory}

\author{Domenico Gazzillo, Riccardo Fantoni \& Achille Giacometti
 \thanks{$^\ast$Corresponding
author. Email: gazzillo@unive.it  \vspace{10pt}
\newline\centerline{\tiny{ {\em Molecular Physics}}}
\newline\centerline{\tiny{ISSN 0026-8976 print/ ISSN
1362-3028
 online
\textcopyright 2005 Taylor \& Francis Ltd}}
\newline\centerline{\tiny{ http://www.tandf.co.uk/journals}}
\newline \centerline{\tiny{DOI:
10.1080/002689700xxxxxxxxxxxx}}}$^\ast$\\\vspace{6pt}Istituto Nazionale 
per la Fisica della Materia and
Dipartimento di Chimica Fisica, Universit\`a di Venezia, S. Marta DD
2137, I-30123 Venezia, Italy}  \received{v3.2 released August 2005}

\begin{document}
\maketitle

\begin{abstract}
\noindent We discuss phase coexistence of polydisperse colloidal suspensions
in the presence of adhesion forces. The combined effect of polydispersity
and Baxter's sticky-hard-sphere (SHS) potential, describing hard spheres
interacting via strong and very short-ranged attractive forces, give rise,
within the Percus-Yevick (PY) approximation, to a system of coupled
quadratic equations which, in general, cannot be solved either analytically
or numerically. We review and compare two recent alternative proposals,
which we have attempted to by-pass this difficulty. In the first one,
truncating the density expansion of the direct correlation functions, we
have considered approximations simpler than the PY one. These $C_{n}$
approximations can be systematically improved. We have been able to provide
a complete analytical description of polydisperse SHS fluids by using the
simplest two orders $C_{0}$ and $C_{1}$, respectively. Such a simplification
comes at the price of a lower accuracy in the phase diagram, but has the
advantage of providing an analytical description of various new phenomena
associated with the onset of polydispersity in phase equilibria (e.g.
fractionation). The second approach is based on a perturbative expansion of
the polydisperse PY solution around its monodisperse counterpart. This
approach provides a sound approximation to the real phase behavior, at the
cost of considering only weak polydispersity. Although a final seattlement
on the soundness of the latter method would require numerical simulations
for the polydisperse Baxter model, we argue that this approach is expected
to keep correctly into account the effects of polydispersity, at least
qualitatively.
\end{abstract}


\section{Introduction}

New technological advances in physico-chemical manipulation of colloidal
mixtures have brought up again the issue of theoretically understanding the
phase behaviour of polydisperse systems \cite{Sollich02}. `Polydispersity'
in colloidal solutions means that, due to their production process,
suspended macroparticles with the same chemical composition cannot be
exactly identical to each other, but in general have different sizes, and
possibly different surface charges, shapes, etc. In practice, a polydisperse
system can be reckoned as a mixture with very large -- or essentially
infinite -- number $M$ of different species or components, identified by one
or several parameters ($M$ large but finite refers to \textit{discrete
polydispersity}, whereas $M\rightarrow \infty $ with a continuous
distribution of polydisperse parameters corresponds to \textit{continuous
polydispersity}). The present paper will consider discrete polydispersity of
spherical colloidal particles, with their diameter being the only
polydisperse attribute (size-polydispersity).

When polydispersity is not negligible, the phase behaviour becomes much
richer, but the determination of phase transition boundaries requires a much
more involved formalism, compared to the monodisperse counterpart. In fact,
the coexistence condition in terms of intensive variables requires that all
phases must have equal temperature, pressure and chemical potentials of the $%
M$ components. In the presence of polydispersity, one should thus solve a
number of equations of the order of $M^{2}$, a task which is practically
impossible for $M$ large or infinite.

However, the study of phase equilibria can conveniently start from the
appropriate thermodynamic potential, which is the Helmholtz free energy $A$
when the experimentally controlled variables are temperature, volume and
numbers of different colloidal species. In the one-component case, the
coexistence condition of equal pressure and chemical potential has a simple
geometrical interpretation in terms of free energy density $a$: the
densities of two coexisting phases are determined by constructing a
double-tangent to $a$ plotted versus particle density. This recipe leads to
the well known Maxwell construction, which connects suitably selected points
along a van der Waals (vdW) subcritical isotherm, in order to 'reduce' its
unphysical loop to a constant-pressure line characteristic of a first-order
phase transition.

In the polydisperse case, a significant progress in the very difficult
problem of predicting phase equilibria can be obtained for models with 
\textit{truncatable} free energies \cite{Sollich02}. Here `truncatable'
means that the excess free energy of the polydisperse system turns out to
depend only on a \textit{finite} number of moments of the distribution
corresponding to the polydisperse attribute (the diameter $\sigma $ in the
simplest cases). For spherical colloids, the excess free energy of the vdW
model extended to polydisperse fluids has such a truncatable structure. Due
to this property, this vdW theory has often been employed as the simplest
model to investigate the effects of polydispersity on the gas-liquid
transition \cite{Sollich02,Bellier-Castella00}. On the other hand, the
influence of polydispersity on freezing has been addressed by using the
hard-sphere (HS) mixture model, which also admits a truncatable free energy 
\cite{Sollich02} (for the fluid phase, the
Boublik-Mansoori-Carnahan-Starling-Leland (BMCSL) \cite{Mansoori71}
expression was employed). It is worth recalling that it is currently believed that
size-polydispersity might destabilize crystallization, eventually
inhibiting freezing above a certain `terminal' value of polydispersity \cite%
{Sollich02}.

The present paper focuses on -- and reviews -- a number of recent attempts
to investigate polydisperse phase equilibria, at least within some
approximations, for another prototype model useful for studying colloidal
suspensions, namely Baxter's sticky-hard-spheres (SHS) \cite{Baxter68}. Here
the particles are hard spheres with a surface adhesion, and the
corresponding potential can be obtained as a limit of an attractive
square-well which becomes infinitely deep and narrow, according to a
particular prescription which ensures a finite non-zero contribution of
adhesion to the second virial coefficient (`sticky limit'). For the
one-component version of this model, Baxter and collaborators \cite{Baxter68}
solved the Ornstein-Zernike (OZ) integral equation coupled with the
Percus-Yevick (PY) approximation (`closure'). This \textit{fully analytical}
solution allows to determine all structural and thermodynamical properties
of the SHS fluid. On the other hand, the multi-component PY solution, which
soon followed Baxter's work \cite{Perram75,Barboy79}, is practically
inapplicable in the presence of significant polydispersity. In fact, it
requires the computation of a set of parameters $\left\{ \Lambda
_{ij}\right\} $ determined by a system of $M(M+1)/2$ quadratic equations,
which -- in general -- cannot be solved even numerically for a mixture with
a large number of components. Moreover, even in the presence of a general
solution for this non-linear algebraic system, the problem of phase
coexistence would still remain out of reach in view of the previous remarks.

In a series of recent papers \cite%
{Gazzillo00,Gazzillo02,Gazzillo03,Fantoni05,Fantoni06,Gazzillo06}, we have
attempted to make some progress along two different lines.

First, starting from the \textit{density} expansion of the cavity function
at contact, we have considered a sequence of simpler approximations
(compared to the PY one) \cite%
{Gazzillo00,Gazzillo02,Gazzillo03,Gazzillo04,Fantoni05}. Within the two
simplest ones among these approximations, denoted as $C_{0}$ and $C_{1}$
(for reasons which will become clear in the following), we have been able to
derive analytically all relevant information regarding structure and
thermodynamics, including the phase coexistence, in view of the fact that
the corresponding free energy turns out to be truncatable \cite{Fantoni05}.
Due to the simplicity of $C_{0}$ and $C_{1}$, it is however reasonable to
expect these approximations to fail at high packing fractions, with a
consequently incomplete or even incorrect description of the effects of
polydispersity on the phase diagram.

Therefore, in collaboration with Peter Sollich, we have recently explored a
second approach \cite{Fantoni06}, where the expansion variable (which must
be small) is an appropriate polydispersity
index. In such a way, we have tried to solve the non-linear algebraic system
-- involved in the PY result -- \textit{perturbatively in polydispersity},
starting from the monodisperse PY solution.

\section{Baxter's SHS model and PY solution}

The SHS model is defined as limiting case of \ a particular square-well (SW)
model \cite{Baxter68}, based upon a potential including steeply repulsive
core and short-ranged attractive tail, i.e.

\begin{equation}
\phi _{ij}^{\mathrm{Baxter\ SW}}\left( r\right) =\left\{ 
\begin{array}{ccc}
+\infty  &  & 0<r<\sigma _{ij}\equiv (\sigma _{i}+\sigma _{j})/2 \\ 
-\epsilon _{ij}^{\mathrm{Baxter~SW}} &  & \sigma _{ij}\leq r\leq
R_{ij}\equiv \sigma _{ij}+w_{ij} \\ 
0 &  & r>R_{ij}\text{ ,}%
\end{array}%
\right.   \label{eq0}
\end{equation}%
with%
\begin{equation}
\epsilon _{ij}^{\mathrm{Baxter~SW}}=k_{B}T\ \ln \left( 1+t_{ij}\frac{\sigma
_{ij}}{w_{ij}}\right) \text{ ,}
\end{equation}%
where $\sigma _{i}$ is the HS diameter of species $i$, $\epsilon _{ij}^{%
\mathrm{Baxter~SW}} >0$ and $w_{ij}$ are the depth and the width of the well,
respectively, $k_{B}$ denotes Boltzmann's constant, $T$ the temperature.
Moreover, 
\begin{equation}
t_{ij}=\frac{1}{12\tau _{ij}}\geq 0,
\end{equation}%
where the conventional Baxter parameter $\tau _{ij}$ is an unspecified
increasing function of $T$, while $\tau _{ij}^{-1}$ measures the strength of
surface adhesion or `stickiness' between particles of species $i$ and $j$.

The `sticky limit' of $\phi _{ij}^{\mathrm{Baxter~SW}}\left( r\right) $
corresponds to taking $w_{ij}\rightarrow 0$. While the SW width goes to
zero, its depth $\epsilon _{ij}^{\mathrm{Baxter~SW}}$ diverges, giving rise
to a Dirac delta function in the Boltzmann factor \cite{Baxter68}, i. e.%
\begin{equation}
e^{-\beta \phi _{ij}^{\mathrm{SHS}}\left( r\right) }=\theta \left( r-\sigma
_{ij}\right) +t_{ij}\ \sigma _{ij}\delta \left( r-\sigma _{ij}\right) 
\label{eq1}
\end{equation}%
where $\beta =\left( k_{B}T\right) ^{-1}$, while $\theta $ and $\delta $ are
the Heaviside step function and the Dirac delta function, respectively.

The advantage of the sticky limit is that one effectively deals with a
single parameter $\tau _{ij}$ for each pair, rather than a combination of
energy and length scales (as occurs in the square-well model, for which no
analytical solution is known). On the one hand, this particular limit has
the disadvantage of introducing some pathologies in the model, notably in
the one-component case \cite{Stell91}. On the other hand, Baxter's model
represents the simplest paradigmatic example of a combination of steep
repulsion and short-range attraction which entails a complete analytical
solution in the one-component case, within a robust approximation such as
the PY closure.

In the multicomponent case, the PY solution of the OZ equation in terms of
Baxter's factor correlation function reads \cite{Perram75,Barboy79}%
\begin{equation}
q_{ij}(r)=\left\{ 
\begin{array}{l}
\frac{1}{2}a_{i}(r-\sigma _{ij})^{2}+(b_{i}+a_{i}\ \sigma _{ij})(r-\sigma
_{ij})+\Lambda _{ij},\qquad (\sigma _{i}-\sigma _{j})/2\leq r\leq \sigma
_{ij} \\ 
\qquad 0,\qquad \qquad \qquad \qquad \text{\ \ \ \ \ \ \ \ \ \ \ \ \ \ \ \ \
\ \ \ \ \ \ \ \ \ \ \ \ \ \ \ \ \ \ \ \ \ \ \ \ \ \ \ elsewhere,}%
\end{array}%
\right.   \label{eq2}
\end{equation}%
where the expressions for the parameters $a_{i}$ and $b_{i}$ may be found in 
\cite{Gazzillo00}, while the quantity 
\begin{equation}
\Lambda _{ij}=t_{ij}\ y_{ij}(\sigma _{ij})\ \sigma _{ij}^{2}\ ,  \label{eq3}
\end{equation}%
which depends on the cavity function at contact $y_{ij}(\sigma _{ij})$, must
be solution of the following system of quadratic equations 
\begin{equation}
\Lambda _{ij}=\alpha _{ij}+\beta _{ij}\sum_{m}x_{m}\left[ \Lambda
_{im}\Lambda _{jm}-\frac{1}{2}\left( \Lambda _{im}\Gamma _{mj}+\Lambda
_{jm}\Gamma _{mi}\right) \right] ~,~~~i,j=1,2,\ldots ,M  \label{eq4}
\end{equation}%
Here, $x_{m}$ is the molar fraction of the $m$-th species ($m=1,\ldots ,M$),
while $\alpha _{ij}=t_{ij}\ y_{ij}^{\mathrm{HS-PY}}(\sigma _{ij})\ \sigma
_{ij}^{2}$, \  $\beta _{ij}=12\rho \ t_{ij}\sigma _{ij}$\ ($\rho $ is the
total number density), and $\Gamma _{ij}=\sigma _{i}\sigma _{j}/(1-\eta )$,
with $\eta $ being the packing fraction \cite{Fantoni06}. The solution of
these equations for $\left\{ \Lambda _{ij}\right\} $ is the real bottleneck
of the multi-component PY result, as mentioned in the Introduction: for
large $M$ (and in particular for $M\rightarrow \infty $ ) this calculation
is next to impossible, neither analytically nor numerically.

As a consequence, although the PY closure is commonly believed to be very
sound for short-range potentials (for one-component SHS fluids this was
confirmed by recent numerical simulations \cite{Miller03}), one has to
conclude that, in the multi-component (polydisperse) case, the PY solution
has a very limited practical usefulness, since its solution scheme cannot be
fully accomplished. This is the reason why other possible routes have been
attempted, as we discuss next.

\section{Simplified closures: the class of $C_{n}$ approximations}

A `closure' is a relationship, added to the OZ equation, between the direct
correlation function $c_{ij}(r)$ and $h_{ij}(r)=g_{ij}(r)-1$ or $\gamma
_{ij}(r)=h_{ij}(r)-c_{ij}(r)$ ($g_{ij}(r)$ being the radial distribution
function) \cite{Barrat03}. 

Let us go back to Baxter's SW model given by Eq. (\ref{eq0}) (i.e. \textit{%
before} the `sticky limit'), and consider the following general class of
`closures' \cite{Gazzillo04} 
\begin{equation}
c_{ij}(r)=\left\{ 
\begin{array}{cc}
-1-\gamma _{ij}\left( r\right)  & 0<r<\sigma _{ij} \\ 
c_{ij}^{\mathrm{shrink}}(r) & \text{ \ \ \ \ }\sigma _{ij}\leq r\leq
R_{ij}\qquad  \\ 
0 & \text{ \ \ \ \ \ \ \ \ \ \ \ \ \ \ }r>R_{ij}\text{ .}\qquad 
\end{array}%
\right.   \label{eq6}
\end{equation}%
The expression $c_{ij}(r)=-1-\gamma _{ij}\left( r\right) $ inside the core ($%
r<\sigma _{ij}$) is exact and dictated by the HS potential. The form outside
the well ($r>R_{ij}$) may then be identified with the PY approximation, 
\begin{equation}
c_{ij}^{\mathrm{PY}}(r)=f_{ij}(r)\left[ 1+\gamma _{ij}\left( r\right) \right]
,
\end{equation}%
since, for Baxter's potential, the Mayer function, $f_{ij}(r)=\exp \left[
-\beta \phi _{ij}(r)\right] -1$, vanishes for $r>R_{ij}$. 

The choice of $c_{ij}^{\mathrm{shrink}}(r)$ inside the well (region which
'shrinks' in the sticky limit) defines one particular closure within the
proposed class. Of course, $c_{ij}^{\mathrm{shrink}}(r)=c_{ij}^{\mathrm{PY}%
}(r)$ corresponds to the PY approximation. On the other hand, when $c_{ij}^{%
\mathrm{shrink}}(r)\neq c_{ij}^{\mathrm{PY}}(r)$, we are in the presence of 
\textit{mixed closures,} which have frequently appeared in the literature 
\cite{Herrera91}. In order to define mixed closures simpler than the PY
approximation, we consider the \textit{density} expansion of the exact
direct correlation function \cite{Barrat03}, and denote as $C_{n}$
approximation a truncation of this series to order $O(\rho ^{n})$. The
simplest two approximations are

\begin{equation}
\begin{array}{ccc}
\begin{array}{c}
c_{ij}^{\mathrm{shrink}}(r)=f_{ij}(r) \\ 
\end{array}
&  & 
\begin{array}{c}
\text{(}C_{0}\text{ closure)} \\ 
\end{array}
\\ 
c_{ij}^{\mathrm{shrink}}(r)=f_{ij}\left( r\right) \ \left[ 1+\left( \
\sum_{k}\rho _{k\ }f_{ik}\ast f_{kj}\right) \left( r\right) \right]  &  & 
\text{(}C_{1}\text{ closure),}%
\end{array}%
\end{equation}%
where $\rho _{k\ }$ is the number density of species $k$, while $\ast $
denotes convolutive integration \cite{Gazzillo04}. 

In the `sticky limit' $R_{ij}\rightarrow \sigma _{ij}^{+}$ the well region
shrinks, but a `memory' of the approximation chosen for $c_{ij}^{\mathrm{%
shrink}}$ remains in the solution of the OZ integral equation. In fact,
although all solutions $q_{ij}(r)$ corresponding to closures belonging to
the class given by Eq. (\ref{eq6}) have the same functional form as the PY
solution -- Eq. (\ref{eq2}) -- , each closure is characterized by its own
approximation to $y_{ij}(\sigma _{ij})$, which is involved in the
expressions of the parameters $a_{i}$, $b_{i}$, $\Lambda _{ij}$. For
instance, the $C_{0}$ and $C_{1}$ approximations correspond to 

\begin{equation}
\begin{array}{ccc}
\begin{array}{c}
y_{ij}(\sigma _{ij})=1 \\ 
\end{array}
&  & 
\begin{array}{c}
\text{(}C_{0}\text{ closure)} \\ 
\end{array}
\\ 
y_{ij}(\sigma _{ij})=1+y_{ij}^{(1)}(\sigma _{ij})\ \eta  &  & \text{(}C_{1}%
\text{ closure),}%
\end{array}%
\end{equation}%
which are, respectively, the zeroth-- and first--order truncations of the
density expansion for the exact cavity function at contact (see Ref. \cite%
{Gazzillo04} for details).

While a brute-force truncation of the above-mentioned density expansions
leads to analytical expressions simple enough to be applied to the
multi-component (polydisperse) case, one should reasonably expect less
accuracy, expecially in the high-density regime. In the one-component case,
we can carefully check this point. 

\begin{figure}[hbtp]
\begin{center}
\vskip0.5cm
\includegraphics[angle=0,width=15cm]{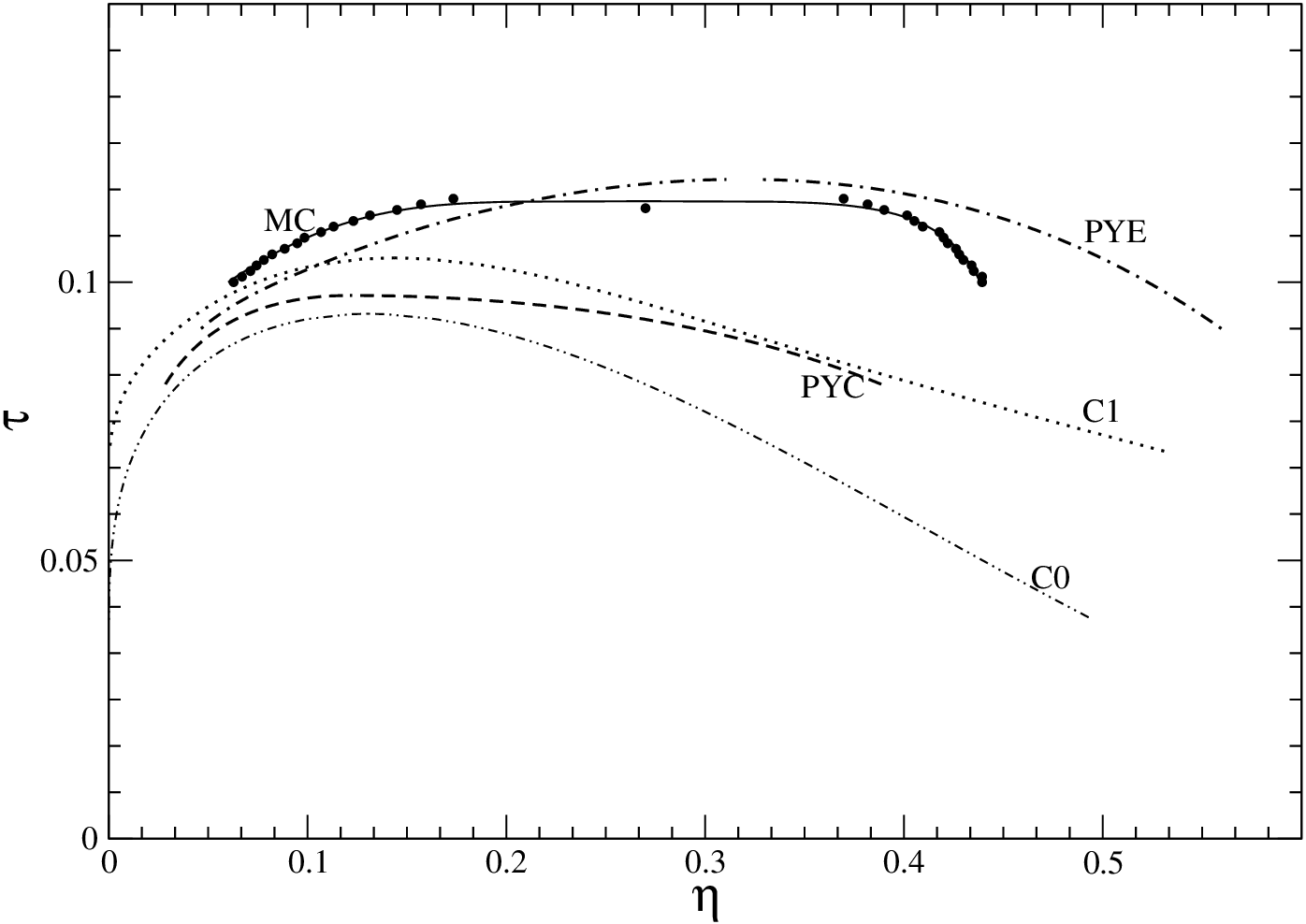} 
\end{center}
\caption{Coexistence (binodal) curves for the one-component Baxter
model. Both compressibility (PYC) and energy (PYE) equation of state
as obtained from the Percus-Yevick approximation (see Ref \cite{Baxter68})
are reported and compared with $C_0$ and $C_1$ approximations from Ref. \cite{Fantoni05}
and with Monte Carlo simulation (MC) from Ref. \cite{Miller03}. In the MC case the
continuous line is simply a guide to the eye.}
\label{fig1}
\end{figure}

In Figure \ref{fig1} coexistence curves obtained from the $C_0$ and $C_{1}$
approximations are compared with the PY ones (using both compressibility and
energy routes), and with Monte Carlo simulations from Ref. \cite{Miller03}.
It is apparent how the PY energy route (PYE) yields a rather precise
representation of the MC results, unlike the compressibility route (PYC). It
is worth noting that the results stemming from the $C_{1}$ approximation,
although rather close to the MC data in the low-density branch, clearly fail
to accurately reproduce them for higher densities, as expected.

In spite of their lack of accuracy, the $C_{0}$ and $C_{1}$ approximations
provide however a rather sound basis for getting some insight into phase
equilibria of polydisperse SHS fluids, since they allow simple analytical,
or semi-analytical, treatments. 

A first important feature of the $C_{0}$ and $C_{1}$
approximations for polydisperse SHS is that the corresponding free energy
has a \textit{truncatable} structure, that is it depends upon
few ($4$ at most) moments of the (discrete) size distribution, $\xi _{\nu
}=(\pi /6)\rho \sum_{j}x_{j}\sigma _{j}^{\nu }$ with $\nu =0,1,2,3$.

A second remarkable fact is that the $C_{0}$ and $C_{1}$ approximations are
able to describe the so-called \textit{fractionation}
phenomena characteristic in phase equilibria of polydisperse systems. While
we refer to a recent review \cite{Sollich02} for detailed description of the
increased complexity in the polydisperse phase diagrams, here we just
mention the two important points. First, fractionation means that daughter
phases, obtained from demixing of a parent homogeneous phase, need not have
the same composition of the parent phase. As a consequence, there is no a
single coexistence line (`binodal') as in the one-component case, but one
rather finds a \textit{cloud curve}, representing the temperature-density
dependence line of the low-density majority phase ('gas'), and a \textit{%
shadow curve} representing the temperature-density dependence of the
high-density minority phase (incipient 'liquid'). While for one-component
systems these two curves are identical, for polydisperse systems in general they are
not, with the exception of the critical point where they
coincide by definition.

However, in order to apply the $C_{0}$ and $C_{1}$ approximations to the
multi-component SHS model, we have to tackle a further important problem,
that is the definition of the stickiness parameters $\tau _{ij}$.

\section{Size-dependence of stickiness parameters}

In mixtures, $\tau _{ij}$ will depend on the particular pair $i,j$
considered, and reasonably should be expected to be related to the particles
sizes. Assuming that we are dealing only with size-polydispersity, we can
always decouple temperature and adhesion as 
\begin{equation}
\frac{1}{\tau _{ij}}=\frac{1}{\tau }\ \epsilon _{ij}=\frac{1}{\tau }\mathcal{%
F}\left( \sigma _{i},\sigma _{j}\right) 
\end{equation}%
where the last equality stems from the assumption of size-polydispersity and
of a purely pairwise potential. Unfortunately, the exact form of the
size-dependence of these stickiness parameters is still an open problem, due
to the lack of experimental and theoretical insights on this \cite{Gazzillo06}.
On the other hand, few guidelines -- based on
arguments discussed in Refs. \cite{Fantoni05,Fantoni06} -- provide, as
reasonable and plausible, the following dependences 
\begin{equation}
\epsilon _{ij}=\mathcal{F}(\sigma _{i},\sigma _{j})=\left\{ 
\begin{array}{ll}
\sigma _{0}^{2}/\sigma _{ij}^{2} & \mbox{Case I}~, \\ 
\sigma _{i}\sigma _{j}/\sigma _{ij}^{2} & \mbox{Case II}~, \\ 
1 & \mbox{Case IV}~, \\ 
\sigma _{0}/\sigma _{ij} & \mbox{Case V}~.%
\end{array}%
\right.   \label{eq9}
\end{equation}%
Here $\sigma _{0}$ is a characteristic reference length (e.g. the parental
mean diameter) and the numbering of the various cases follows the
convention of previous work  \cite{Fantoni05,Fantoni06}.

\begin{figure}[hbtp]
\begin{center}
\vskip0.5cm
\includegraphics[angle=0,width=15cm]{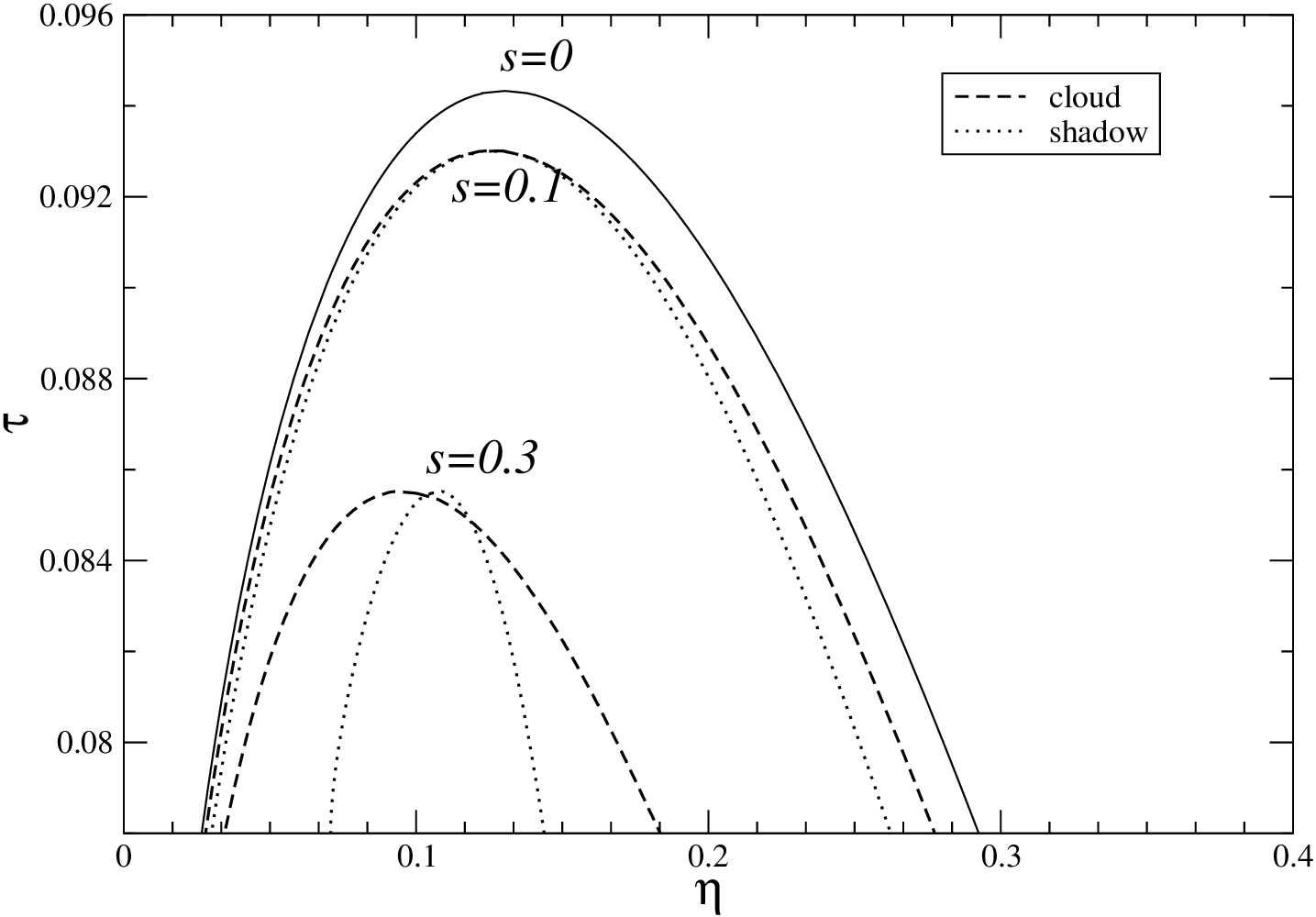} \\
\vskip2.0cm
\includegraphics[angle=0,width=15cm]{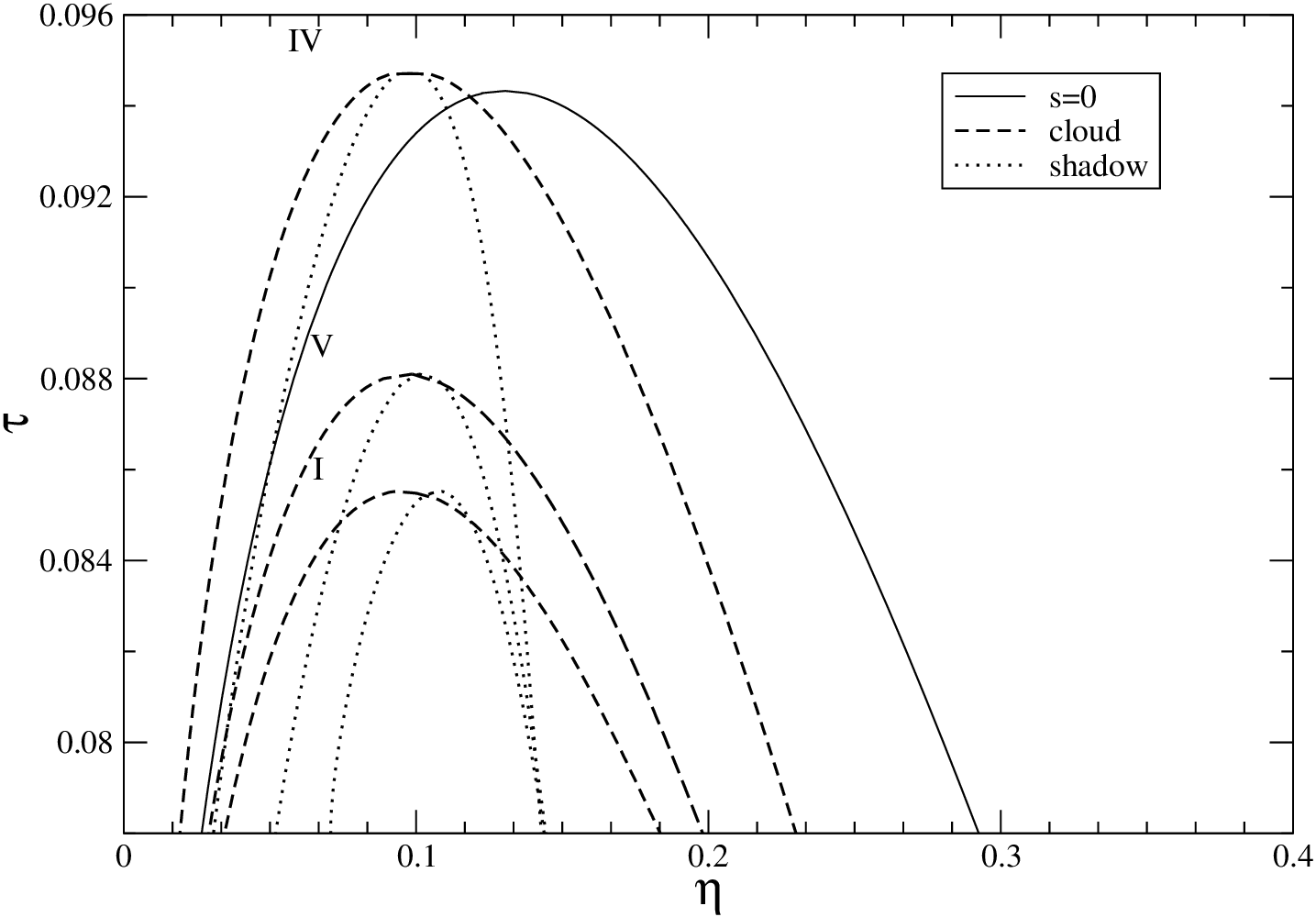} 
\end{center}
\caption{(Top) Cloud and shadow curve for model I within the $C_0$ approximation
at increasing polydispersity: $s=0$, $s=0.1$ and $s=0.3$ . (Bottom) Same as above
for fixed value of polydispersity $s=0.3$ and different choice of the stickiness 
adhesion (model I, IV and V)
 }
\label{fig2}
\end{figure}

Figure \ref{fig2} reports the results of the calculation of the cloud and
shadow curves for polydisperse SHS within the simple $C_{0}$ approximation.
Here and below, polydispersity is measured by an index $s$, which is the
normalized standard deviation of the size distribution. Hence $s=0$ corresponds
to a mono-disperse case, whereas $s=0.1$ and $s=0.3$ indicate moderate and
significant polydispersity, respectively. The top panel of Fig. \ref{fig2}
depicts the results for case I of size-dependence of stickiness parameters.
As $s$ increases, the coexistence region shrinks, thus suggesting that
polydispersity \textit{disfavours} the phase transition. On the other hand,
this trend is markedly case-dependent, as illustrated in the bottom panel of
Fig. \ref{fig2}, where the cloud-shadow pairs with polydispersity $s=0.3$
are displayed for different cases of size-dependence. It can be clearly seen
that while, for cases I and V the same trend is observed, case IV seems to
suggest a \textit{widening} of the phase coexistence region (and hence a
favouring of the phase transition).

In view of the lack of numerical simulation for polydisperse SHS to compare
with, we have no way, at the present stage, to check how realistic those
results are. On the other hand, we might suspect, based on the comparison in
the one-component case, $C_{0}$ to fail to provide accurate representation
in the region of low temperatures and high densities. This is the reason why
other possible approaches have been recently tested. We now illustrate a
different perturbative approach, which has proved to be promising in this
context.

\section{Perturbative treatment of the SHS-PY solution}

The main difficulty in dealing with the PY solution for polydisperse SHS
stems from the solution of the coupled quadratic system of equations (\ref%
{eq4}), as remarked. As the one-component case has a well-defined solution,
one might then suspect that -- for weak polydispersity -- a perturbative
expansion around this solution might include the main effects of
polydispersity. This is in fact what happens, as recently shown \cite%
{Fantoni06} by exploiting a general perturbation theory due to Evans \cite%
{Evans01}. The main idea is that, for weak polydispersity,
size-distributions are narrowly peaked around a mean reference value ($%
\sigma _{0}$ in the present case), and hence all quantities such as%
\begin{equation}
\delta _{i}=\frac{\sigma _{i}-\sigma _{0}}{\sigma _{0}}\ll 1\text{,}
\label{eq10}
\end{equation}%
are small. Therefore one might expand both $\epsilon _{ij},$ and all
quantities appearing in $\Lambda _{ij},$ in powers of $\delta _{i}$. Similar
expansion can be performed in the free energy, and hence all thermodynamic
quantities can be computed. The entire procedure is described in detail in
Refs. \cite{Sollich02,Fantoni06,Evans01}. Here we just summarize the main results.

\begin{figure}[hbtp]
\begin{center}
\vskip0.5cm
\includegraphics[angle=0,height=9.0cm,width=7.5cm]{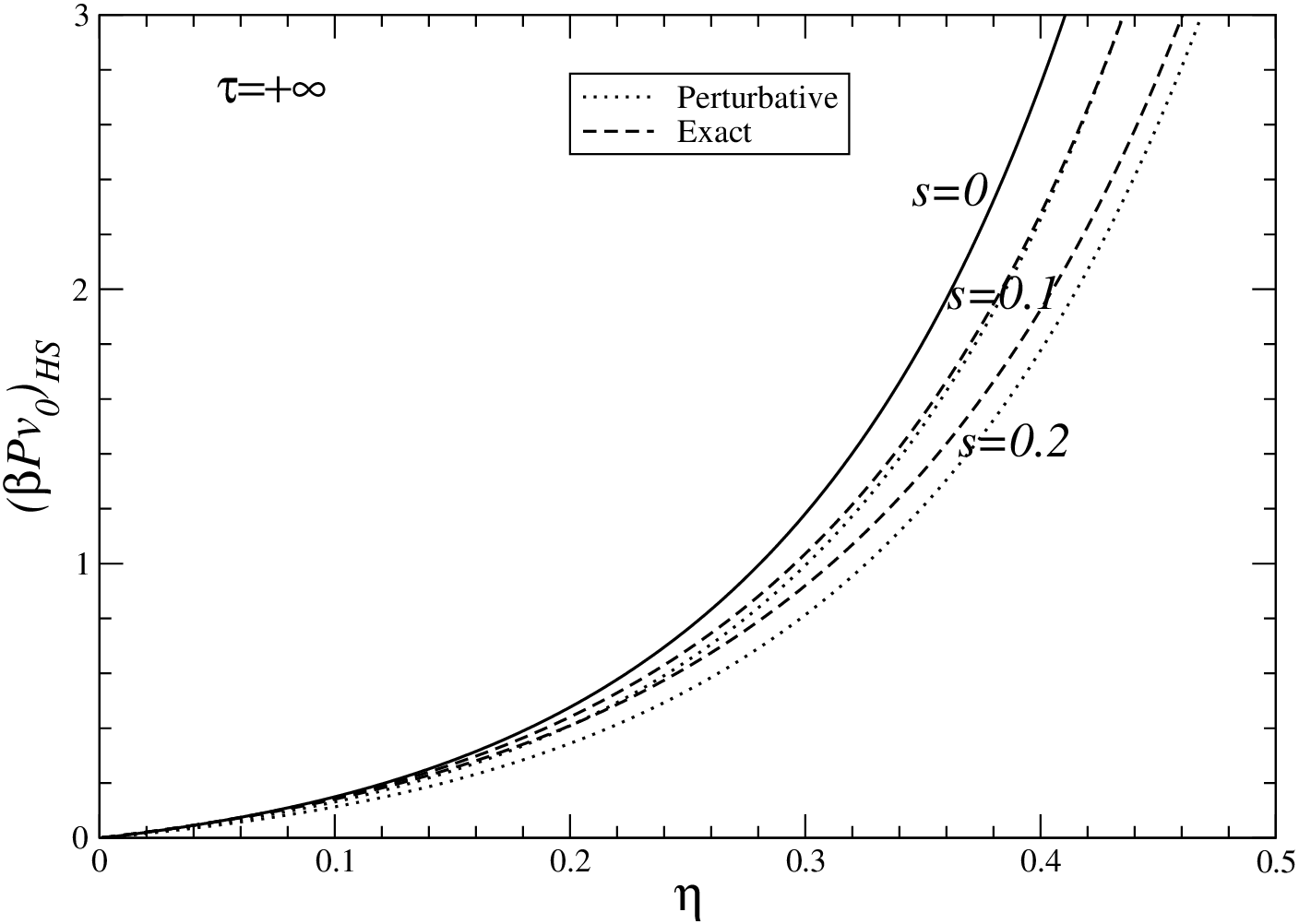}
\includegraphics[angle=0,height=9.0cm,width=7.5cm]{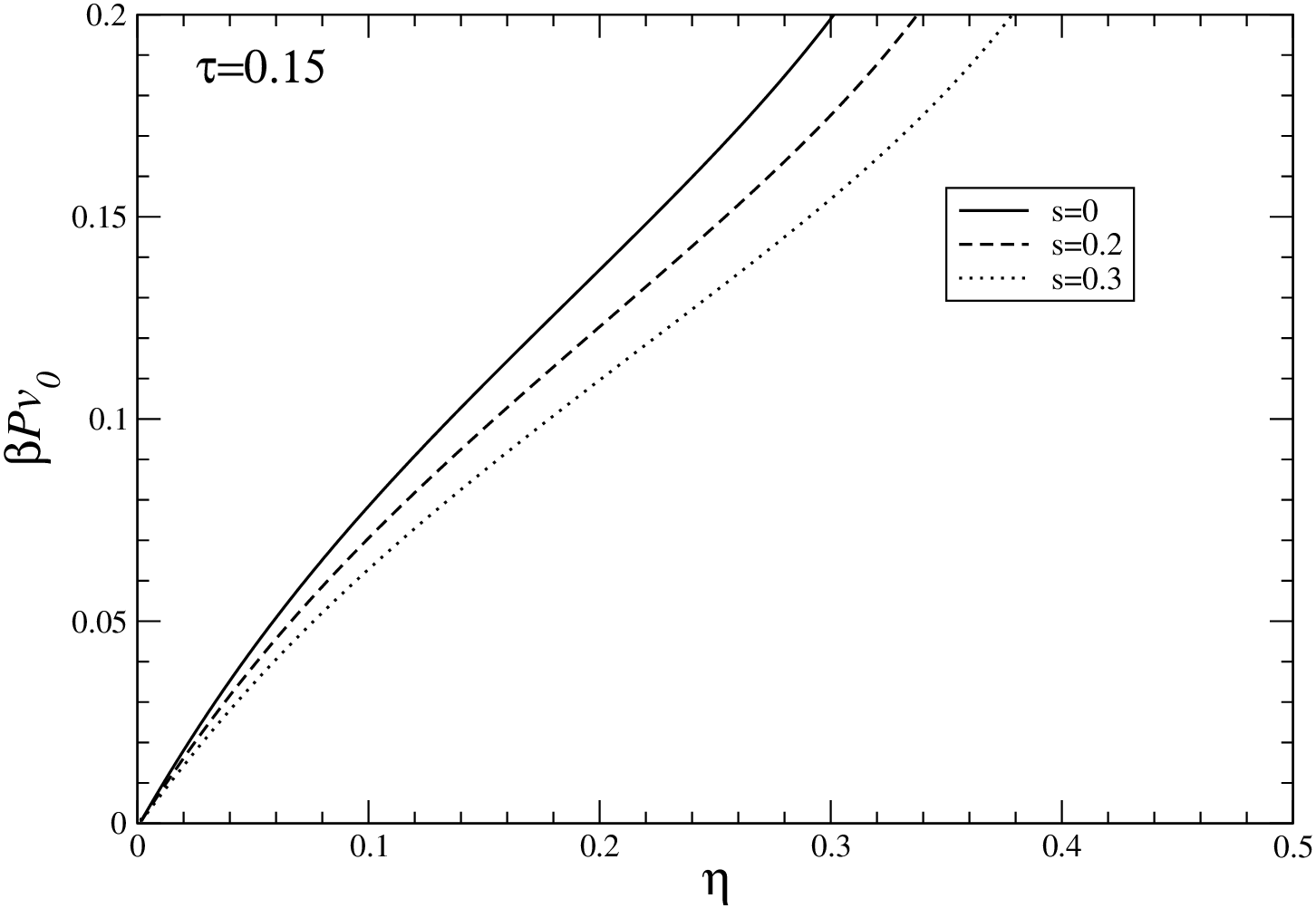} \\
\vskip1.5cm
\includegraphics[angle=0,height=9.0cm,width=7.5cm]{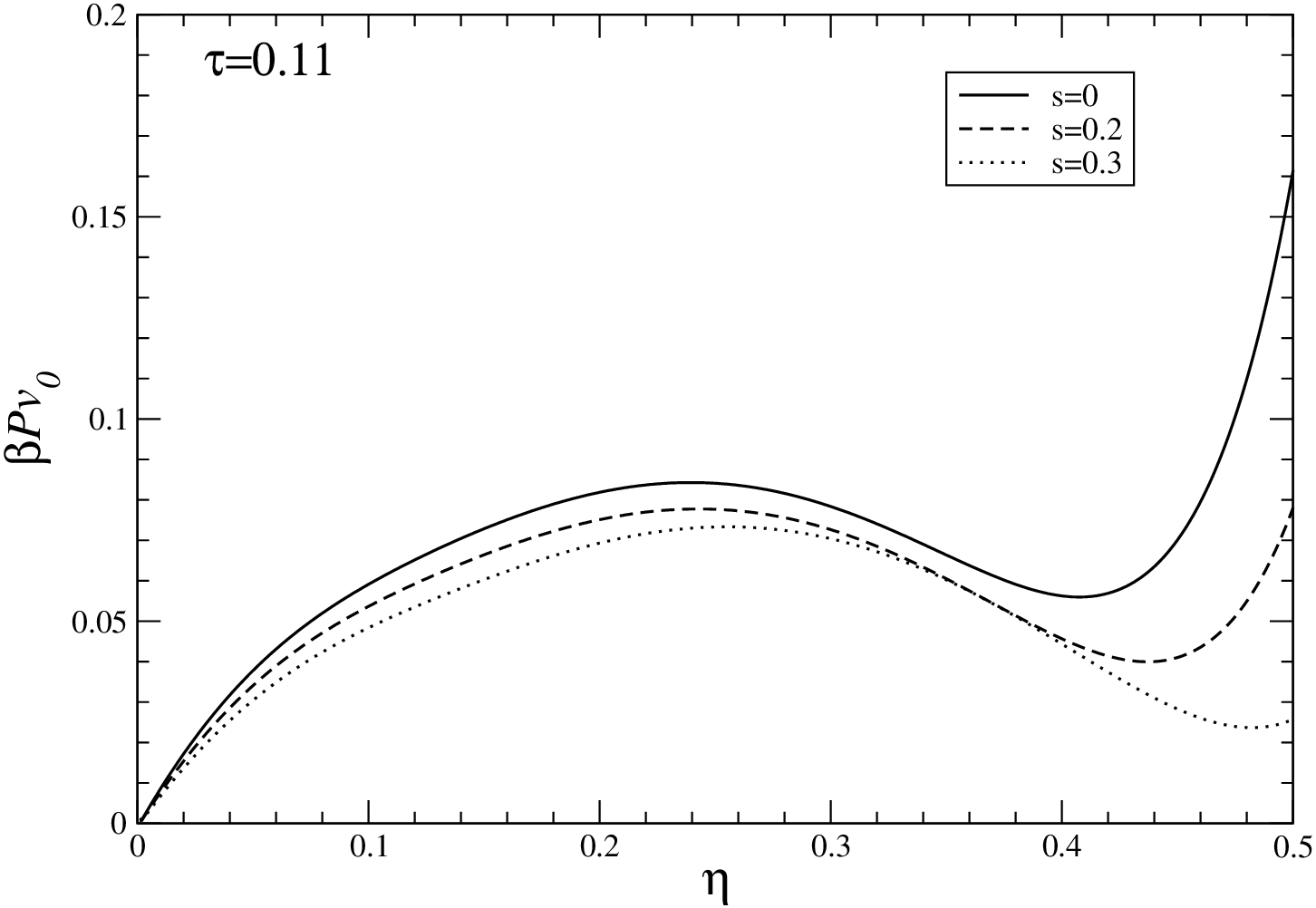} 
\includegraphics[angle=0,height=9.0cm,width=7.5cm]{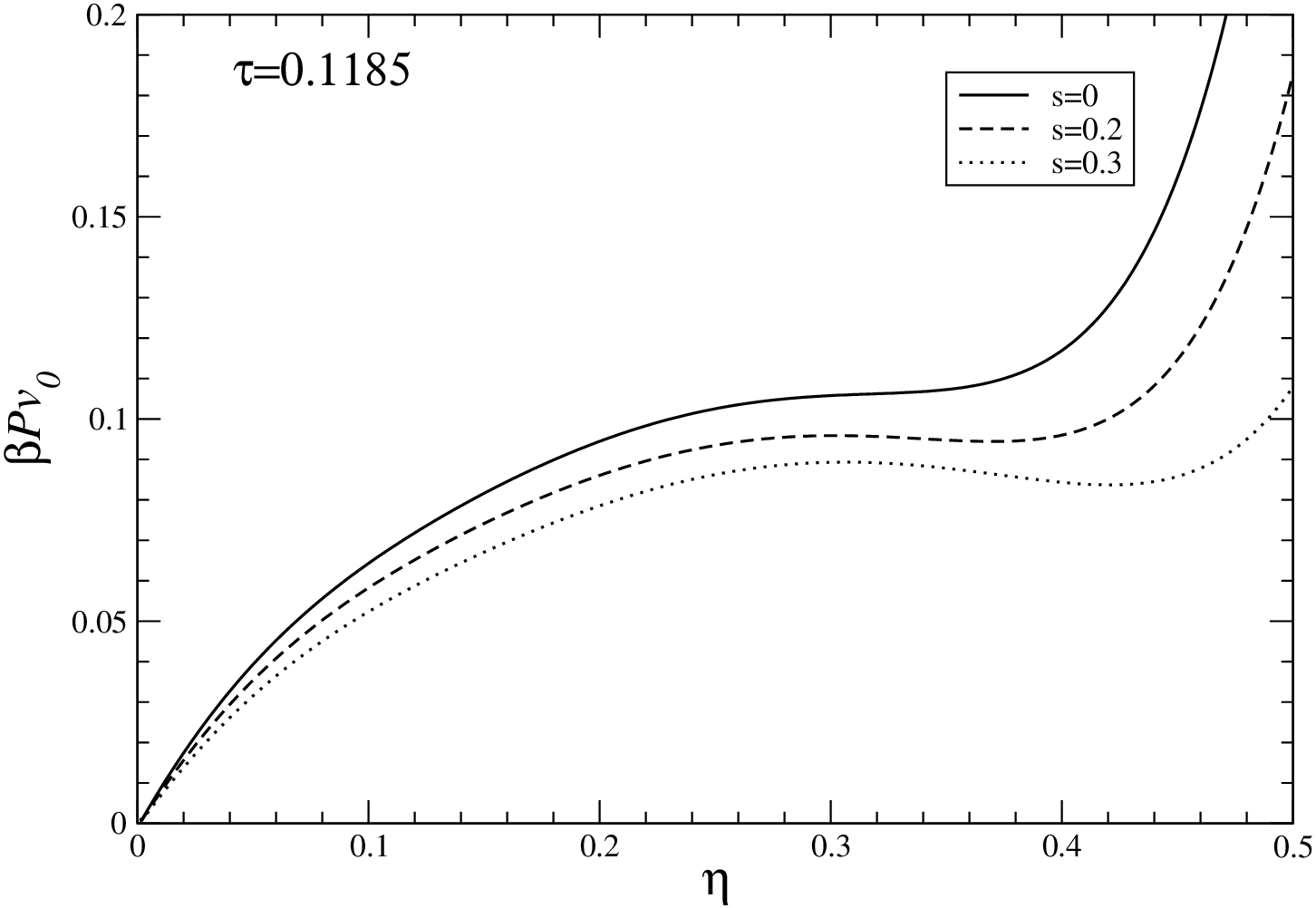}
\end{center}
\caption{Behavior of the energy equation of state 
within our perturbative scheme. In all cases the quantity $\beta P v_0$ is
plotted against the packing fraction $\eta$. In clockwise order, the first curve
(left, top) reports a comparison of the perturbative versus the `exact' BMCSL solution
in the equation of state for polydisperse HS ($\tau=+\infty$). The other curves report 
the perturbative solution
for the energy equation of state within the PY approximation for SHS Baxter model.
Results are depicted for three values of temperature $\tau=0.15> \tau_c$, $\tau=0.1185\sim
\tau_c$ and $\tau=0.1<\tau_c$ and for different degrees of polydispersity. The choice
for the size-dependence of stickiness parameters corresponds to model IV.
 }
\label{fig3}
\end{figure}

The approximate range of validity of the perturbation expansion can be envisaged by considering
the polydisperse HS case where the `exact' BMCSL approximation \cite{Mansoori71} 
can be compared with the corresponding
perturbative solution based on the one-component ($s=0$) counterpart. This is reported
in the top-left part of Figure \ref{fig3}, where the quantity 
$\beta P v_0$ ($v_0=\pi \sigma_0^3/6$)
is plotted against the packing fraction $\eta$ for increasing values of polydispersity. 
It is apparent how the perturbative solution remains rather close to the 'exact'  polydisperse
BMCSL solution even for moderate polydispersity $s \le 0.3$, which is the maximum value
considered in the remaining part of the work.
The  remaining plots in
Figure \ref{fig3} display the effect of polydispersity on the PY pressure equation of state
as obtained from the energy route and for decreasing values of the temperature $\tau$. 
In the one component case $s=0$, a van der Waals loop
starts to appear when we cross the critical temperature $\tau_c \sim 0.1185$ coming
from the high $\tau$ regime. Obviously this signals the onset of a liquid-gas phase
transition, and the corresponding phase diagram can be obtained by a standard
Maxwell construction by connecting appropriate points with the same pressure. 
In the presence of polydispersity (here represented by choice IV for
size-dependence of the stickiness parameters) 
the same procedure {\it cannot} be applied due to fractionation,
as already discussed. Nevertheless, we can clearly see that as $s$ increases, 
the van der Waals loop
region (when present) expands, thus suggesting that phase transition is favored by
the presence of polydispersity.  A similar feature occurs for the polydisperse van der Waals 
model \cite{Bellier-Castella00} and for the numerical results of the PY compressibility 
equation of 
state \cite{Robertus89} (note that in the latter a {\it gap} rather that a {\it loop} 
is signalling the onset of the transition).
A somewhat surprising feature is that, at fixed packing fraction $\eta$, the pressure decreases
with increasing polydispersity {\it less} in the presence of adhesion rather than 
in its absence (i.e. for the HS case).
An intuitive plausible interpretation of this feature can be found in Ref. \cite{Fantoni06}.

\begin{figure}[hbtp]
\begin{center}
\vskip0.5cm
\includegraphics[angle=0,height=9.0cm,width=7.5cm]{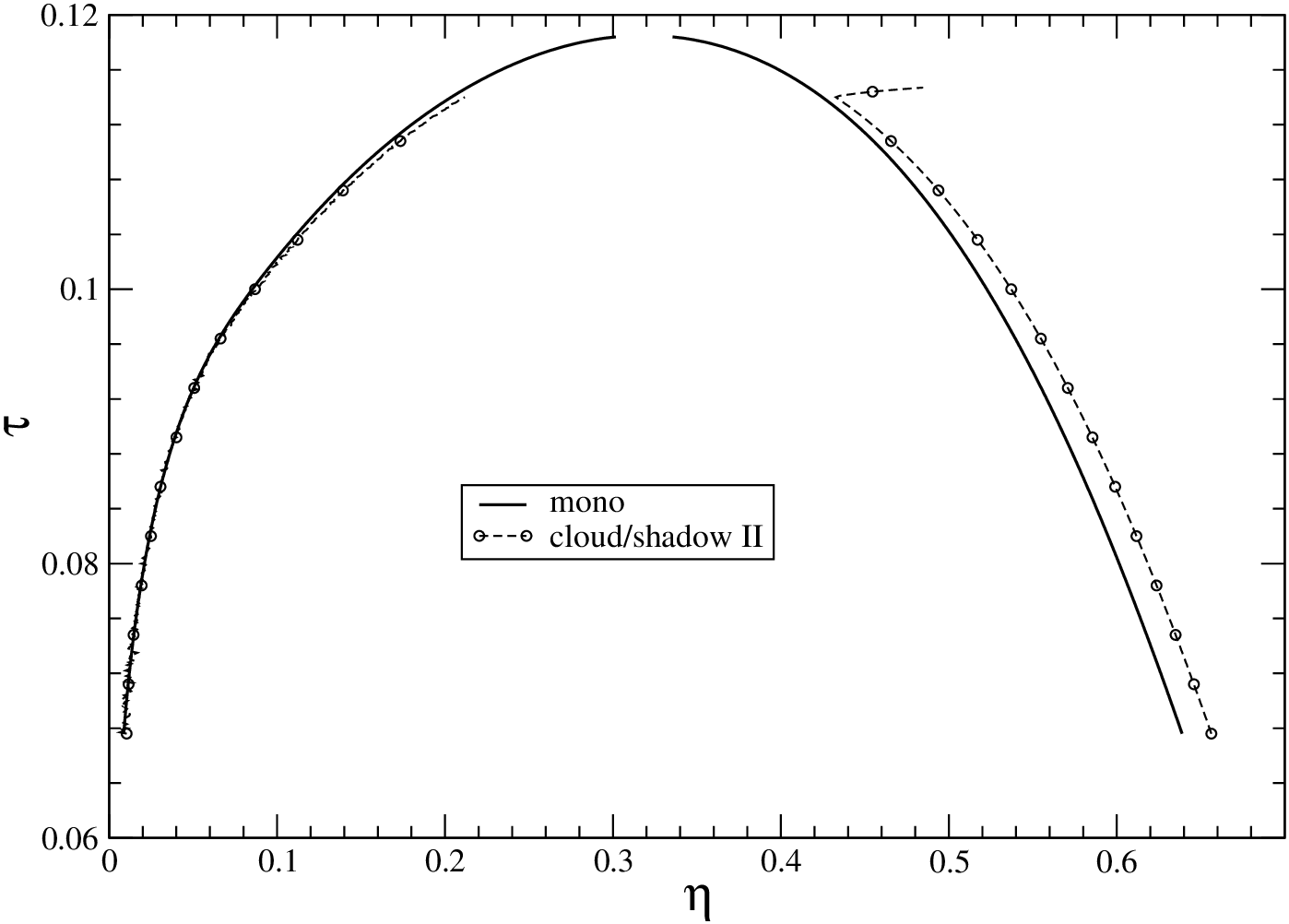} \hskip0.5cm
\includegraphics[angle=0,height=9.0cm,width=7.5cm]{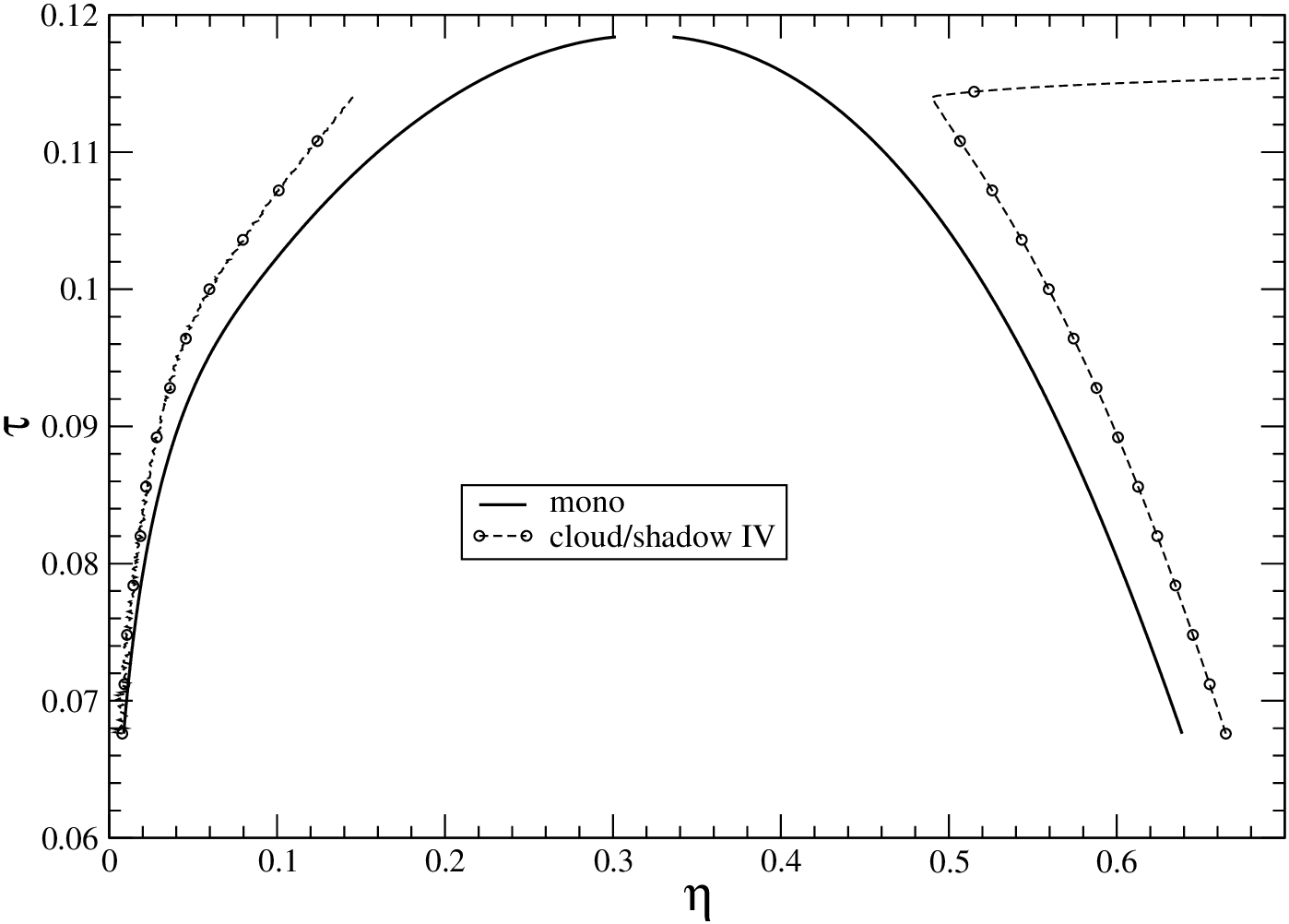} \\
\vskip1.5cm
\includegraphics[angle=0,height=9.0cm,width=7.5cm]{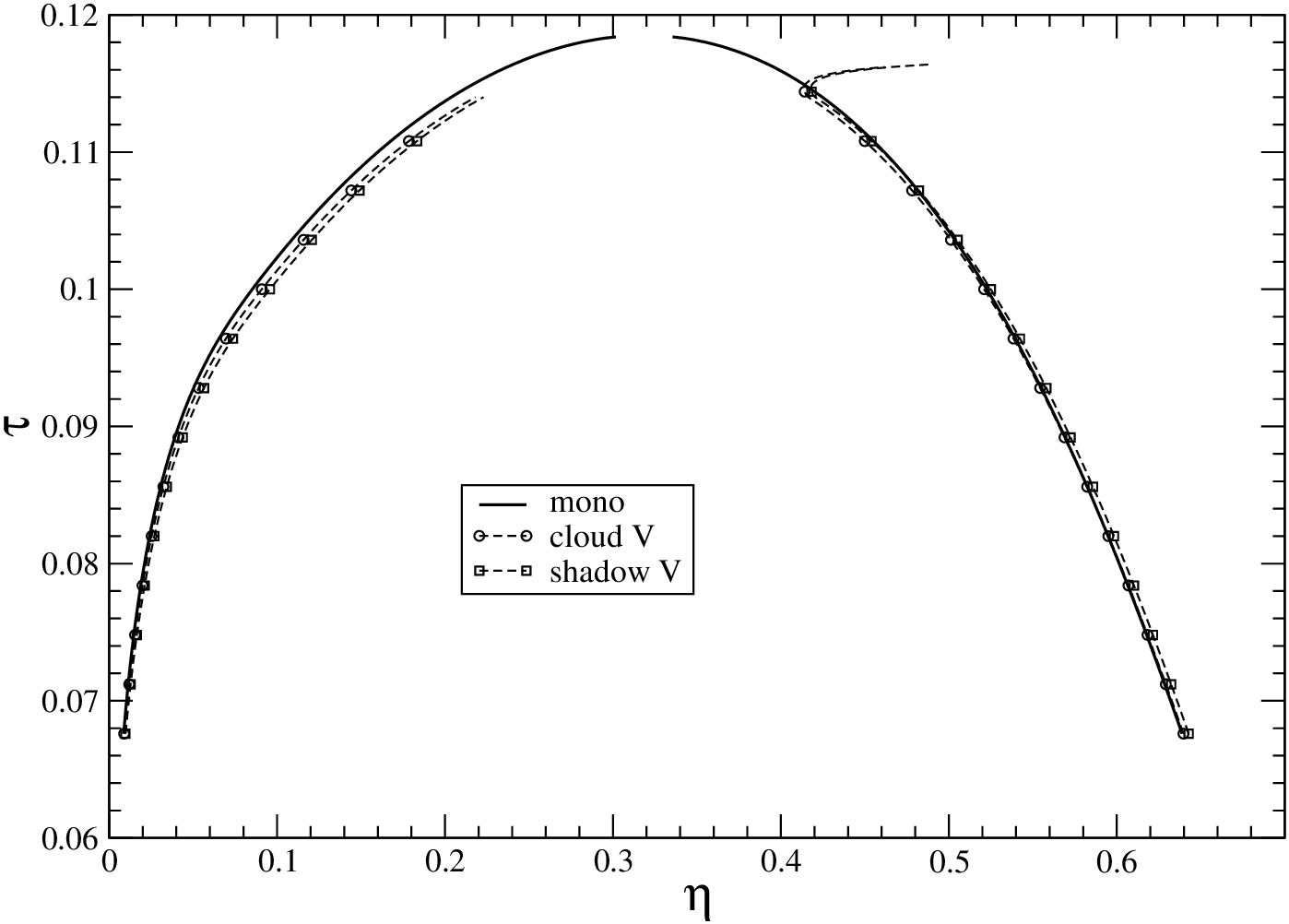} \hskip0.5cm
\includegraphics[angle=0,height=9.0cm,width=7.5cm]{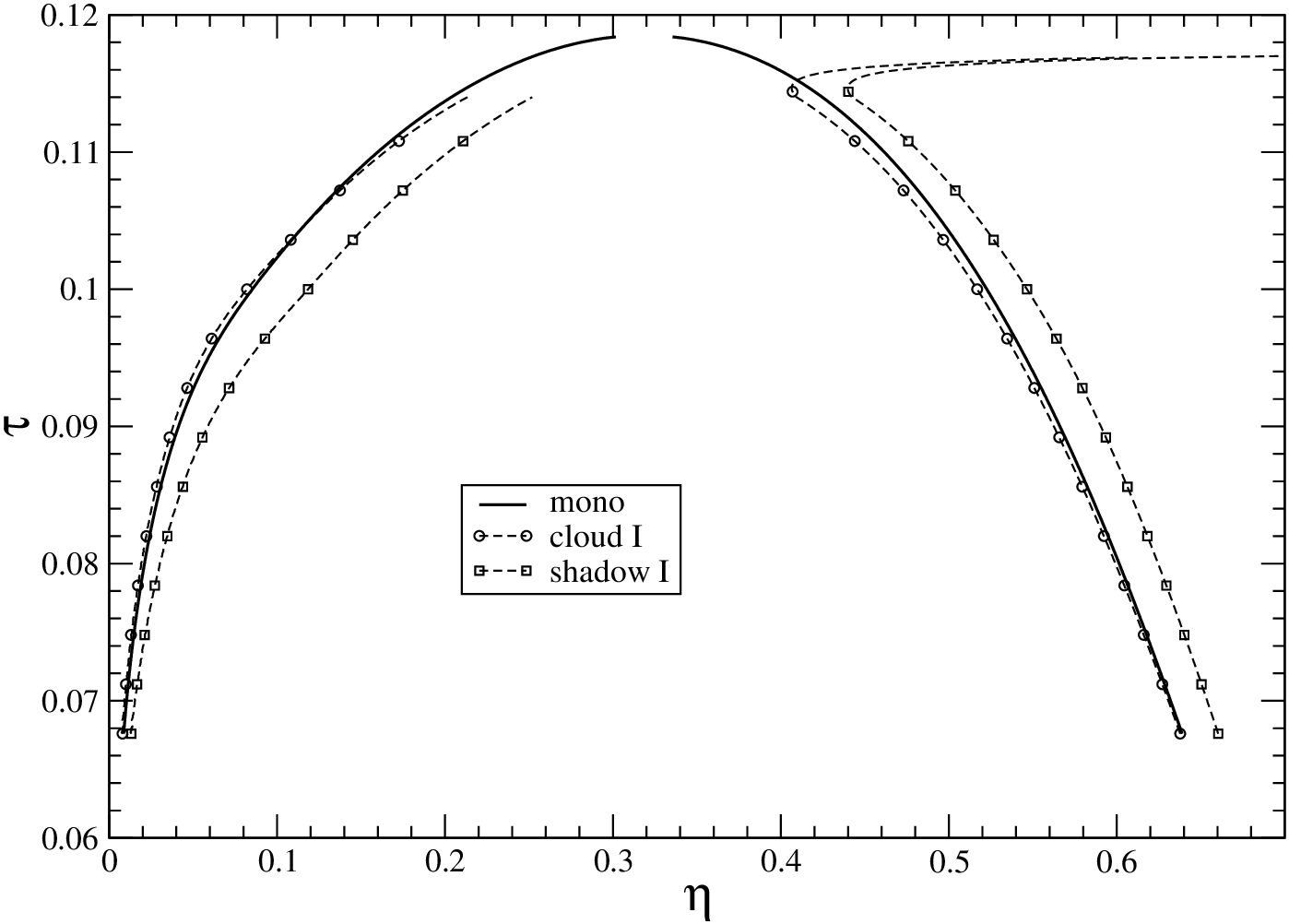}
\end{center}
\caption{Cloud/Shadow pairs from the perturbative results for the PY solution of SHS Baxter model.
In clockwise order the results for choices II, IV (top) and I,V (bottom) are depicted. 
In the two top panels, the cloud
and shadow curves coincide to this order in perturbation, whereas in the bottom panels they 
are different.
In order to have all pictures on the same scale, the selected value for polydispersity is $s=0.3$ 
for models II, IV (top) and $s=0.1$ for models I, V (bottom). 
In all cases the continuous curve represents
the monodisperse ($s=0$) result. 
 }
\label{fig4}
\end{figure}

The same perturbative approach allows the determination of the cloud and
shadow curves for the various cases of size-dependence of $\ \tau _{ij}^{-1}$%
. This is reported in Figure \ref{fig4} for cases II, IV (top panel) and I,
V (bottom panel). In the first case the cloud and shadow lines collapse into
a single curve, and this can be understood on the basis of the particular
scaling properties of the free energy to this order in perturbation theory 
\cite{Fantoni06}. In all cases there is a breakdown of the perturbation
theory on approaching the critical point, and this is a known general
drawback of Evans' perturbative scheme. Nevertheless, in all cases and to
this order in perturbation theory, there is a tendency of the phase
coexistence region to \textit{increase} with polydispersity, in qualitative
agreement with the intuitive picture obtained from Figure \ref{fig3}.

It is worth stressing the difference with previous non-perturbative results
stemming from the $C_{0}$ solution, where all different cases (with the
notable exception of IV) were predicting a reduction of the phase
coexistence region. While in the $C_{0}$ description we have provided a
careful treatment of polydispersity at the expenses of accuracy of the
exploited approximation, in the perturbative description of the PY solution,
polydispersity is assumed to be small and hence one might suspect that
solutions with large polydispersities cannot fit within this picture. On a
balance, nevertheless, we would favour the latter rather than the former
description. An almost correct representation of the one-component
counterpart, is a necessary requirement to check the effect of
polydispersity on it, and we are not aware of any physical or experimental
system where the effects of polydispersity are so strong that they could not
kept into account, at least at the simplest qualitative level, by the
perturbative scheme proposed here. Along this line, some further proposals
have been put forward in Ref. \cite{Fantoni06} to derive a phenomenological
BMCSL-like approximation for SHS, which might be regarded as our 'best and
simplest guess' to the \textit{exact} phase behavior of polydisperse SHS.
Even on the size-dependence of $\tau _{ij}^{-1}$ some possible support of
the proposed forms may be argued \cite{Fantoni06,Gazzillo06}.

\section{Conclusions}

In this work we have summarized recent advances in predicting theoretically the
phase diagram for polydisperse suspensions of colloidal particles with
surface adhesion, within the simple description of Baxter's model. Emphasis
was put on the crucial -- unsolved -- step required to get the
multi-component SHS-PY solution, and the proposed recipies to deal with this
problem. This first one is based on a simplification of the closure. It has
the advantage of allowing a complete analytical analysis on the effects of
polydispersity, including fractionation, but has the disadvantage of a very
questionable accuracy. The second one is based on a perturbative method,
starting from the energy PY one-component solution, which is known to
provide an accurate description of the phase diagram. The drawback of this
scheme is that it works for mild polydispersity, and that cannot describe
the changes of the critical point region. Notwithstanding these limitations,
this novel approach is expected to find practical application in the
interpretation of all those phenomena where Baxter's model and
polydispersity both play a privileged role.

\begin{flushleft}
\textbf{Ackowledgments} \\
Part of the work appearing here has been obtained in
collaboration with Peter Sollich.\\
\end{flushleft}


\newpage



\end{document}